\documentclass[aps,prl,superscriptaddress,twocolumn,longbibliography]{revtex4-2}
\usepackage{bbm}
\usepackage{graphicx}
\usepackage{dcolumn}
\usepackage{bm}
\usepackage{subfigure}
\usepackage{amsmath}
\usepackage{amssymb}
\usepackage{feynmf}
\usepackage{hyperref}
\usepackage{color}

\usepackage{attachfile}

\usepackage{times}

\begin{document}

\title{Sign-reversal  Anomalous Hall effect driven by a magnetic transition in Cr$_{7-\delta}$Te$_8$}

\author{Bowen Chen}
\affiliation{Center for Neutron Science and Technology, Guangdong Provincial Key Laboratory of Magnetoelectric Physics and Devices, School of Physics, Sun Yat-Sen University, Guangzhou, 510275, China}
\affiliation{State Key Laboratory of Optoelectronic Materials and Technologies, Sun Yat-Sen University, Guangzhou, Guangdong 510275, China}

\author{Xiaokai Wu}
\affiliation{Center for Neutron Science and Technology, Guangdong Provincial Key Laboratory of Magnetoelectric Physics and Devices, School of Physics, Sun Yat-Sen University, Guangzhou, 510275, China}
\affiliation{State Key Laboratory of Optoelectronic Materials and Technologies, Sun Yat-Sen University, Guangzhou, Guangdong 510275, China}

\author{Zhiyu Liao}
\affiliation{Beijing National Laboratory for Condensed Matter Physics, Institute of Physics, Chinese Academy of Sciences, Beijing 100190, China}
\affiliation{School of Physical Sciences, University of Chinese Academy of Sciences, Beijing 100049, China}

\author{Zhendong Fu}
\affiliation{Neutron Science Platform, Songshan Lake Materials Laboratory, Dongguan, Guangdong 523808, China}

\author{Bing Xu}
\email{Corresponding author: bingxu@iphy.ac.cn}
\affiliation{Beijing National Laboratory for Condensed Matter Physics, Institute of Physics, Chinese Academy of Sciences, Beijing 100190, China}
\affiliation{School of Physical Sciences, University of Chinese Academy of Sciences, Beijing 100049, China}

\author{Meng Wang}
\email{Corresponding author: wangmeng5@mail.sysu.edu.cn}
\affiliation{Center for Neutron Science and Technology, Guangdong Provincial Key Laboratory of Magnetoelectric Physics and Devices, School of Physics, Sun Yat-Sen University, Guangzhou, 510275, China}

\author{Bing Shen}
\email{Corresponding author: shenbing@mail.sysu.edu.cn}
\affiliation{Center for Neutron Science and Technology, Guangdong Provincial Key Laboratory of Magnetoelectric Physics and Devices, School of Physics, Sun Yat-Sen University, Guangzhou, 510275, China}
\affiliation{State Key Laboratory of Optoelectronic Materials and Technologies, Sun Yat-Sen University, Guangzhou, Guangdong 510275, China}

\begin{abstract}
The search for exotic spin configurations and related novel transport properties continues to be fueled by the promise of new electronic states and outstanding candidate components for spintronic applications. In layered  Cr$_{7-\delta}$Te$_8$, the applied field drives a before unreported magnetic transition revealed by the alternating current magnetic susceptibility measurements around room temperature. This observed magnetic transition results in a sign change for the anomalous Hall effect which exhibits non-monotonous temperature dependence. 
The prominent topological Hall effect (THE) with a large value of 1$\mu \Omega \cdot cm$ has been observed without breaking the inversion symmetry for Cr$_{7-\delta}$Te$_8$. This robust THE can persist up to room temperature attributed to the nonzero fluctuation-driven scalar spin chirality. The complicated interactions of long-range and short-range magnetic orders lead to rich exotic magnetic states with related novel transport properties in Cr$_{7-\delta}$Te$_8$.

\end{abstract}
\pacs{}
\date{\today}
\maketitle
\section{I. INTRODUCTION}
Topologically nontrivial magnetic and electronic structures always attract extraordinary attention in condensed matter physics\cite{tokura2019magnetic_topological,narang2021topology_topological2}. The non-collinear magnetic orders in various layered magnets are considered to host highly nontrivial physics with exotic phenomena. For example, the  skyrmion\cite{ndiaye2017topological_skyrmion1,leroux2018skyrmion2}, a kind of topological-protected magnetic texture, usually emerges by breaking the inversion symmetry due to Dzyaloshinskii-Moriya interaction (DMI)\cite{seshadri2018topological_DMI1,fernandez2019symmetry_DMI2}. Its motion driven by the applied field can result in an addition of Hall contribution called the topological Hall effect (THE) \cite{ndiaye2017topological_skyrmion1,leroux2018skyrmion2}. With further investigations, it is found that this exotic transport property can also emerge in some centrosymmetric magnetic systems such as the van der Waals (vdW) two-dimensional (2D) ferromagnetic(FM) Fe$_3$GeTe$_2$\cite{zhang2018emergence_Fe3GeTe2}, FM Cr$_2$Ge$_2$Te$_6$\cite{carteaux1995crystallographic_1Cr2Ge2Te6,zhang2015robust_2Cr2Ge2Te6} and antiferromagnetic (AFM) MnBi$_2$Te$_4$\cite{zhang2019topological_1MnBi2Te4,deng2020quantum_2MnBi2Te4}. The recent experimental and theoretical studies revealed the contributions of short-range(SR) magnetic orders in some centrosymmetric systems prompting the realization of peculiar quantum states at the temperatures far above the long-range(LR) AFM or FM transitions\cite{EuZn2As2_short_range_order, EuCd2As2_spin_fluctuation, Tm166_kagome}.   The interactions of LR and SR magnetic orders can host rich exotic nontrivial topological electronic or magnetic states and related effects \cite{EuZn2As2_short_range_order, EuCd2As2_spin_fluctuation, Tm166_kagome, EuSn2As2_variation, nisoli2013colloquium_SRandLR1, mydosh2015spin_SRandLR2}. These results enlighten us to explore high-temperature nontrivial topological and exotic magnetic phenomena in materials with large magnetic fluctuations and disorders.

Chromium-based chalcogenides such as Cr$_m$Te$_n$ can offer tunable structural phases and their magnetic properties are sensitive to stoichiometric variations\cite{CrTe_phase_structure,hashimoto1969magnetic_Cr7Te8_magneticastructure,akram1983magnetic_tunnable_structure,dijkstra1989band_calculation}. These compounds consist of alternating stacks of Cr-deficient and Cr-full layers along the $c$ direction strongly affecting the magnetic structures\cite{wang2019magnetic_1Cr5Te8, Magnetic_anisotropy_AHE_MCr5Te8}.CrTe$_2$, a van der Waals (vdW) material, exhibits intrinsic ferromagnetism with a Curie temperature (\textit{T$_C$}) up to 300 K and the large atomic magnetic moment of 0.21$\mu_B$/Cr  promising 2D ferromagnet for exotic low-dimensional 
spintronic applications\cite{zhang2021room_CrTe2}. With increasing the Cr concentration for Cr$_m$Te$_n$ such as in Cr$_5$Te$_8$, it is predicted that a geometrically frustrated structure with strong magnetic fluctuations emerges resulting in a nontrivial magnetic structure and related transport properties such as the THE and large anomalous Hall effect (AHE)\cite{wang2019magnetic_1Cr5Te8,tang2022phase_2Cr5Te8}. These exotic phenomena resulting from magnetic ordering and spin-orbital coupling (SOC), are conducive to the construction of next-generation non-dissipating electronic devices. With further increasing the Cr 
concentration for Cr$_m$Te$_n$ such as in Cr$_{0.87}$Te, nanoscale Bloch-type magnetic bubbles with random chirality emerge due to the competition between a magnetic dipole interaction and uniaxial easy axis anisotropy \cite{liu2022magnetic_Cr7Te8_skyrmion}. These rich magnetic effects make Cr$_m$Te$_n$  an intriguing material platform to explore nontrivial magnetic states and phenomena. Especially, it is noticed that vacancies in this system would easily change the ratio of Te and Cr which is key to the crystal and magnetic structure\cite{CrTe_phase_structure,akram1983magnetic_tunnable_structure,hashimoto1969magnetic_Cr7Te8_magneticastructure}. The related new phases and interactions of various magnetic orders are still far beyond understanding. In this paper, we performed systematic magneto-transport and magnetic investigations on the single crystals of Cr$_{7-\delta}$Te$_8$. Interestingly, in this centrosymmetric system, the prominent THE with a positive value emerges in a large temperature region and persists even above room temperature. The applied field drives a before unreported magnetic phase transition around 280 K. The competition of various magnetic states leads to a sign change and the non-monotonous temperature dependence of AHE. The results reveal the rich and complicated magnetic interactions among LR and SR magnetic orders promising the Cr$_{7-\delta}$Te$_8$ as a good candidate material for new-generation high-temperature spintronic devices. 

\begin{figure}
    \centering
    \includegraphics[width=3.4in]{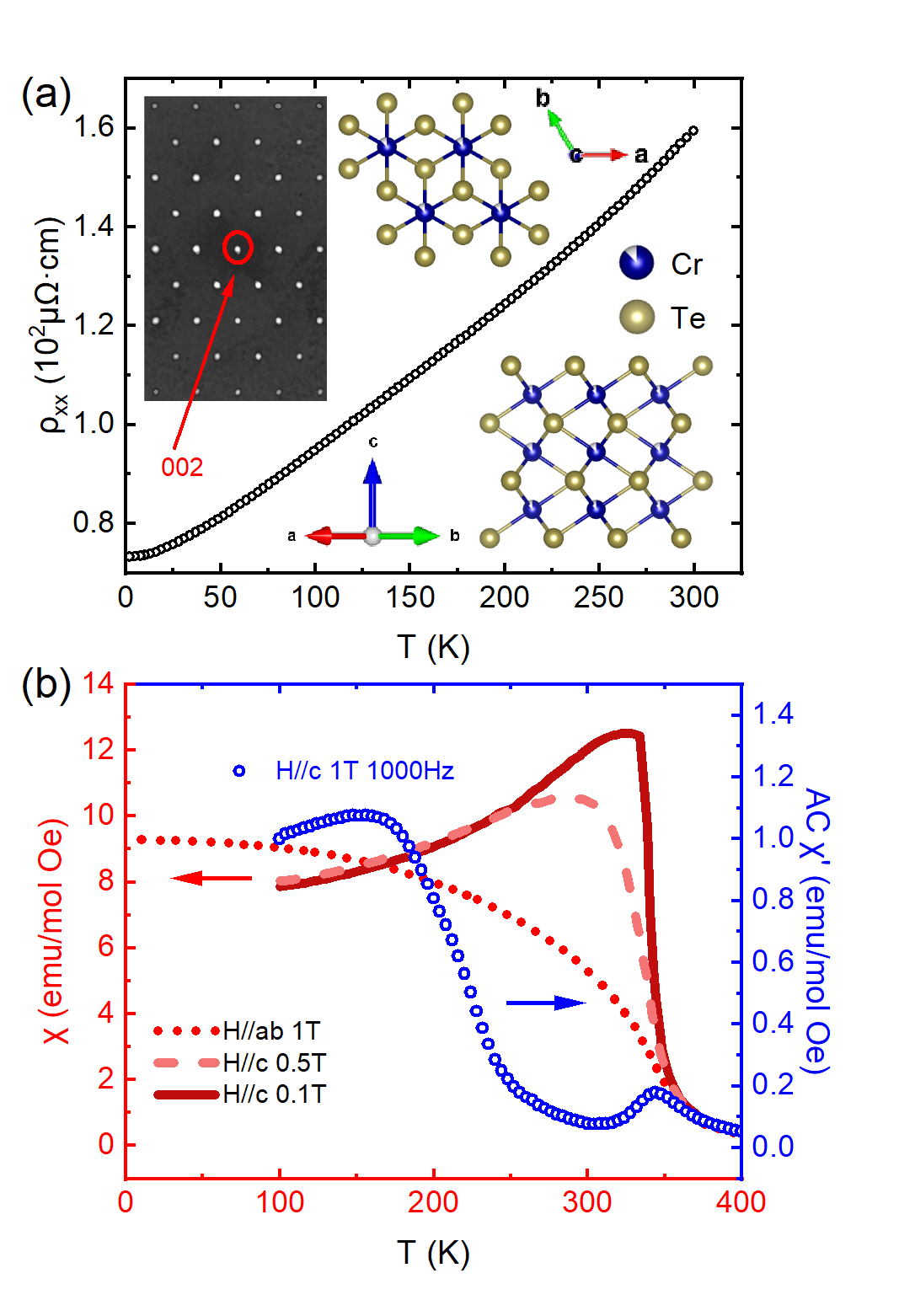}
    \caption{ (a) Temperature-dependent resistivity of Cr$_{7-\delta}$Te$_8$. The left upper inset is the single crystal diffraction pattern in the reciprocal space and the top view of the schematic of the crystal structure of Cr$_{7-\delta}$Te$_8$. The right lower inset is the side view of the schematic of the crystal structure of Cr$_{7-\delta}$Te$_8$.  (b) Temperature-dependent direct current magnetic susceptibility (left red) and alternating current magnetic susceptibility of Cr$_{7-\delta}$Te$_8$ (right blue).  }
    \label{fig:FIG. 1.}
\end{figure}

\section{II. EXPERIMENTAL DETAILS}
Single crystals of Cr$_{7-\delta}$Te$_8$ were grown by the chemical vapor transport method.  The CrTe powder and agent TeCl$_4$ were sealed into an evacuated quartz tube. Then, the quartz tube was placed into a tube furnace, and the high-temperature end was kept at 900 $^\circ$C for 14 days. The Structure and elemental composition of crystals were confirmed by single crystal x-ray diffraction (XRD) and energy dispersive x-ray spectroscopy (EDX) respectively.  Electrical transport measurements were performed on a commercial physical property measurement system (PPMS Dynacool, Quantum Design). The direct current (DC) magnetization measurements and alternating current (AC) susceptibility measurements were carried on the vibrating sample magnetometer (VSM) module and AC measurement system (ACMS) module of the PPMS respectively. 

\begin{figure}
    \centering
    \includegraphics[width=3.4in]{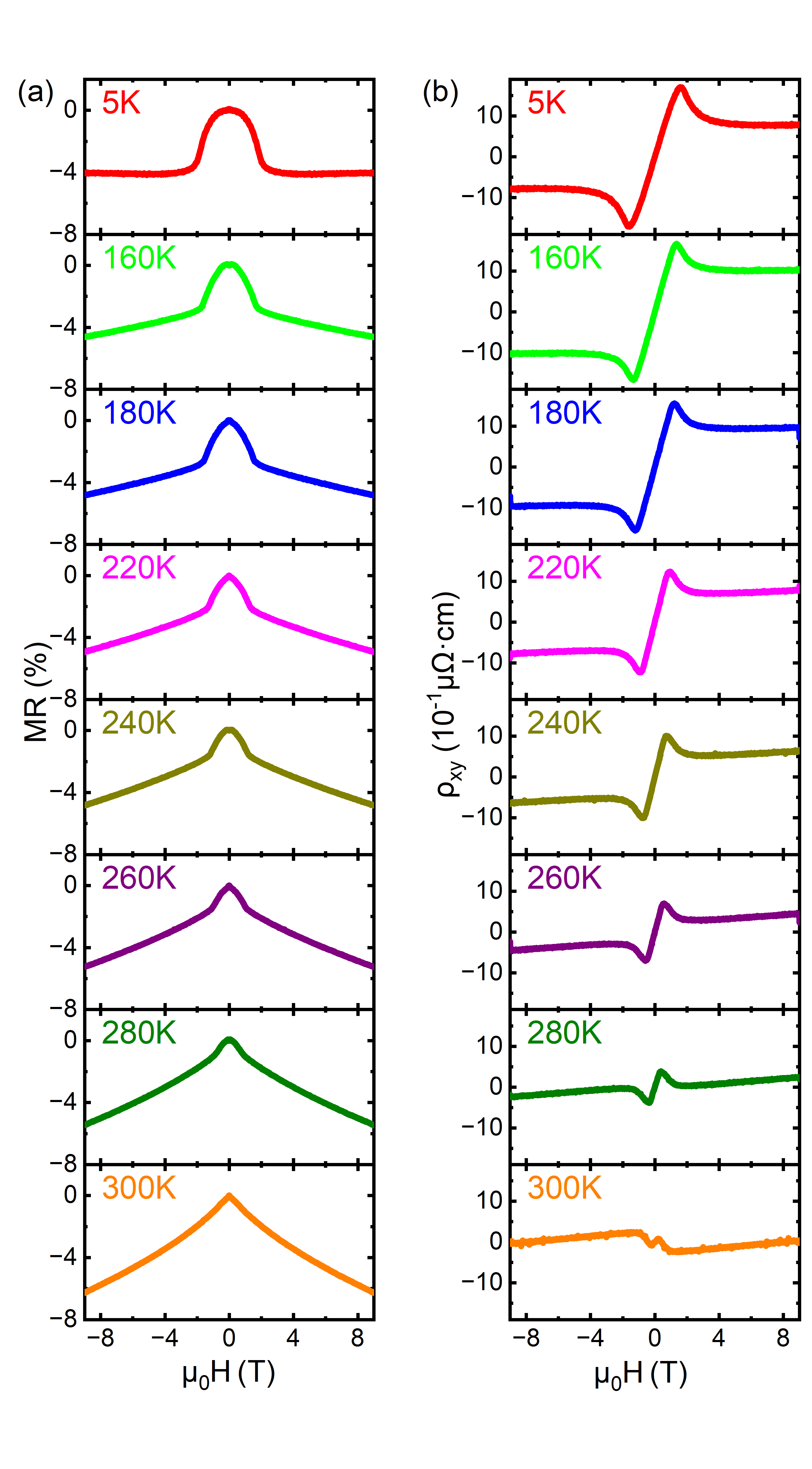}
    \caption{ (a)  Magnetic-field-dependent magneto-resistivity and (b) Hall resistivity of Cr$_{7-\delta}$Te$_8$ at 5 K, 160 K, 180 K, 220 K, 240 K, 260 K, 280 K and 300 K, respectively, with the applied field along the \textit{c} axis of the crystal. }
    \label{fig:FIG. 2.}
\end{figure}

\section{III. RESULTS AND DISCUSSIONS}
\begin{figure*}
    \centering
    \includegraphics[width=6.8in]{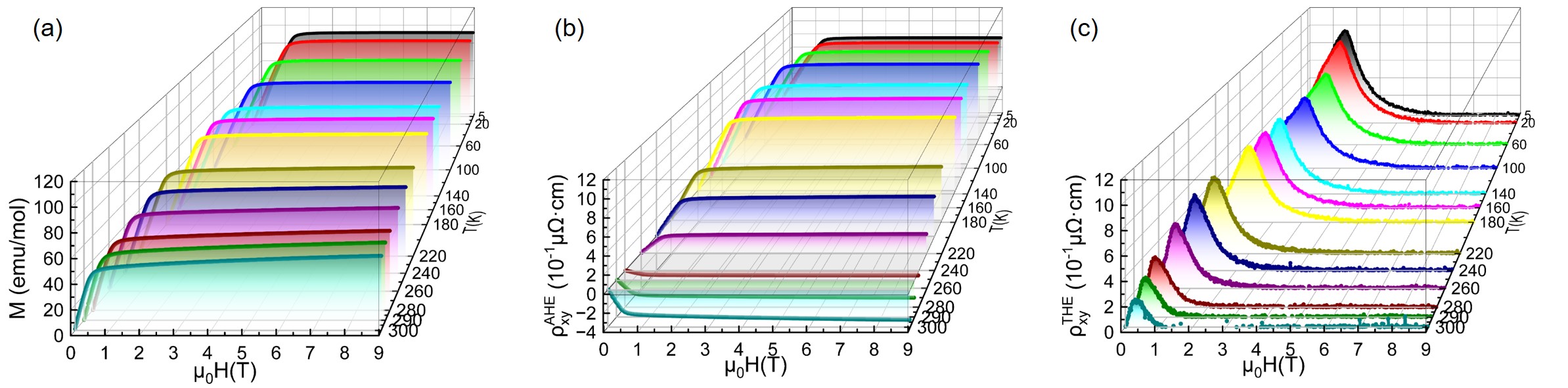}
    \caption{ Magnetic-field-dependent magnetization M($\mu_0$H) (a), anomalous Hall resistivity $\rho_{xy}^{AHE}$ (b) and topological Hall resistivity $\rho_{xy}^{THE}$ (c) of Cr$_{7-\delta}$Te$_8$ at 5 K, 20 K, 60 K, 100 K, 140 K, 160 K, 180 K, 220 K, 240 K, 260 K, 280 K, 290 K and 300 K respectively, with the applied field along the \textit{c} axis of the crystal.}
    \label{fig:FIG. 3.}
\end{figure*}

The temperature-dependent longitudinal resistivity of Cr$_{7-\delta}$Te$_8$ decreases with lowering the temperature exhibiting a typical metallic behavior shown in Fig. 1(a). By the single crystal X-ray diffraction measurement, the crystal structure of Cr$_{7-\delta}$Te$_8$ is revealed shown in the inset of Fig. 1(a). It shares the same structure as CrTe with random Cr vacancies. As shown in Fig. 1(b), the magnetization measurements with the applied field along the \textit{c} axis reveal two magnetic transitions: the high-temperature FM transition around 350 K (\textit{T$_{HF}$}) and the low-temperature transition around 15 K (\textit{T$_{LF}$}) due to the crystal field of Cr. With applying the field parallel to the \textit{ab} plane, the low-temperature transition is absent suggesting the anisotropic magnetic correlations. With increasing the applied field along the \textit{c} axis, the high-temperature FM  transition becomes broad suggesting the magnetic domains establishing. Interestingly, the AC susceptibility measurements reveal that this high-temperature FM transition breaks into two transitions shown in Fig. 1(b). Besides the FM transition around 350 K,  another transition emerges around 280 K  which may be probably ascribed to the transformation of various magnetic states due to the interactions of magnetic fluctuations and  LR magnetic orders (discuss below).

Field-dependent magneto-resistivity (MR defined as MR$=(\rho_{xx}(\mu_0H)-\rho_{xx}(0))/\rho_{xx}(0)\times 100\%$) and Hall resistivity($\rho_{xy}$($\mu_0$\textit{H})) at various temperatures are presented in Fig. 2 (a) and (b) respectively.  At 5 K, clear negative MR 
is observed around $\mu_0 H$= 0 T. With increasing the applied field, the system undergoes a transition around \textit{$\mu$$_0$H$_s$} accompanied by the saturated magnetic moment and almost invisible positive MR (\textit{$\mu_0H$} \textgreater $\mu$$_0$$H_s$ ).  

\begin{figure*}
    \centering
    \includegraphics[width=6.8in]{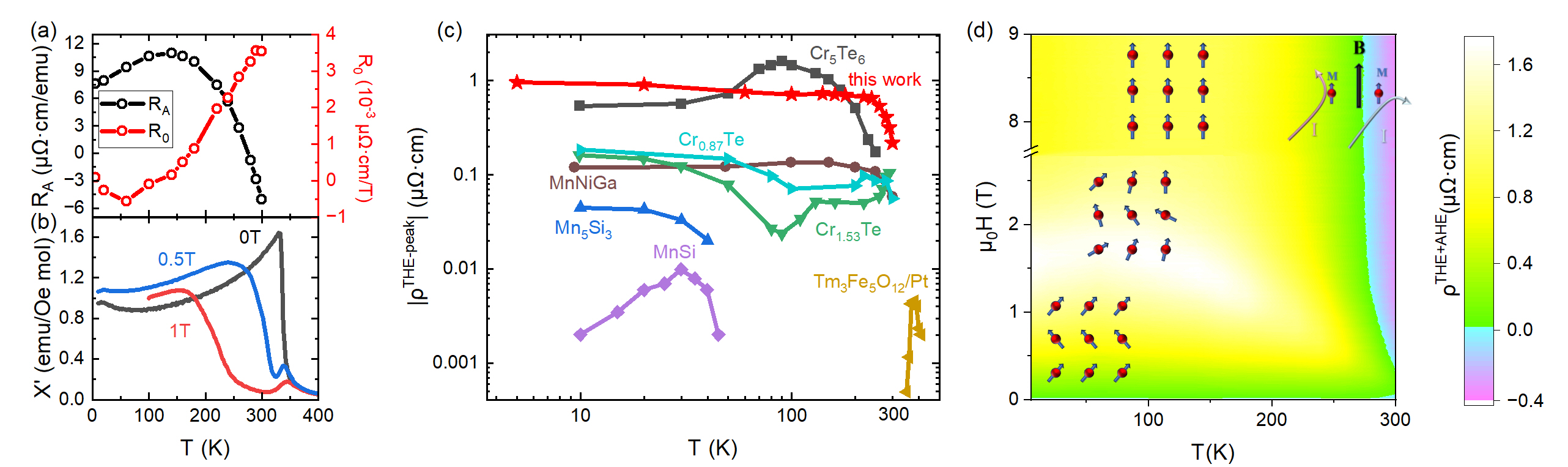}
    \caption{ (a) Temperature-dependent anomalous Hall coefficient (black line) and ordinary Hall coefficient (red line) of Cr$_{7-\delta}$Te$_8$. (b) Temperature-dependent alternating current magnetic susceptibility of Cr$_{7-\delta}$Te$_8$ with various applied fields. (c) The comparison of the topological Hall resistivity for various materials. \cite{Cr1.53_topological_Hall,Cr5Te6_thin_film,liu2022magnetic_Cr7Te8_skyrmion,Mn5Si3,MnNiGa_topological_Hall,MnSi_topologicla_Hall,TmFeO/Pt_topologicla_Hall}(d)The phase diagram of Cr$_{7-\delta}$Te$_8$. The color plot indicates the  total Hall resistivity of AHE and THE.}   
    \label{fig:FIG. 4.}
\end{figure*}

With increasing the temperature, this \textit{$\mu_0H_s$} shifts to lower fields, and the obvious negative MR is observed with \textit{$\mu_0H$} \textgreater \textit{$\mu_0H_s$}. The $\rho$$_{xy}$($\mu$$_0$H) exhibits a strong nonlinear field response. At 5 K,  an obvious abnormal peak emerges in the low field region. With \textit{$\mu_0H$} \textgreater $\mu_0H_s$ $\rho$$_{xy}$($\mu$$_0$$H$) exhibits invisible dependence of the applied field consisting with the saturated magnetic moment.  With increasing the temperature,  a sign change of $\rho$$_{xy}$($\mu$$_0$$H$) emerges in the high-temperature region. In a magnetic system, besides the ordinary Hall contribution (OHE), the magnetization usually gives the anomalous Hall contribution(AHE). For Cr$_{7-\delta}$Te$_8$, after subtracting the ordinary Hall contribution due to the Lorentz force with the linear field response, a discrepancy between $\rho$$_{xy}$($\mu$$_0$$H$) and the magnetization $M$($\mu$$_0$$H$) is observed, suggesting the emergence of an additional Hall contribution (THE) besides AHE. Thus, the Hall resistivity of Cr$_{7-\delta}$Te$_8$ can be expressed as\cite{wang2019magnetic_1Cr5Te8,liu2022magnetic_Cr7Te8_skyrmion}:
\begin{eqnarray}
\rho_{xy}&=\rho_{xy}^{OHE}+\rho_{xy}^{AHE}+\rho_{xy}^{THE}\\
&=R_{0}\mu_{0}H+R_{A}M+\rho_{xy}^{THE},
\end{eqnarray}
where $\rho_{xy}^{OHE}$, $\rho_{xy}^{AHE}$ and $\rho_{xy}^{THE}$ denote the ordinary Hall resistivity, anomalous Hall resistivity, and topological Hall resistivity respectively. And the $R_{0}$ and $R_{A}$ are the ordinary Hall coefficient and the anomalous Hall coefficient, respectively. To further analyze the Hall behavior for Cr$_{7-\delta}$Te$_8$,  $\rho_{xy}^{AHE}$($\mu$$_0$$H$) and $\rho_{xy}^{THE}$($\mu$$_0$$H$) are separated according to the former formula and presented in Fig.3.

In contrast to the $M$($\mu$$_0$$H$) exhibiting  the monotonous temperature dependence, $\rho_{xy}^{AHE}$($\mu$$_0$$H$) behaves a complicated temperature response. With increasing the temperature, it increases to a maximum value, then decreases and eventually changes the sign around room temperature.  The prominent THE with a typical peak in $\rho_{xy}$ emerges in the whole temperature region and keeps a positive value to room temperature. To further analyze these phenomena, temperature-dependent $R_{0}$ and $R_{A}$ are acquired according to the formula shown in Fig. 4(a).  It is observed with increasing the temperature, $R_{A}$ starts to drop around 150 K and changes to a negative value around 280 K. In contrast, the peak of topological Hall resistivity  ($\rho_{xy}^{THE-peak}$ ) exhibits different temperature dependence. These results suggest that AHE and THE are independent of each other in our Cr$_{7-\delta}$Te$_8$ unlike those in the other report before\cite{liu2022magnetic_Cr7Te8_skyrmion}. To understand these phenomena,  we need to consider the detailed magnetic structure and interactions in Cr-Te system. In a perfect CrTe (no Cr vacancies), the magnetic structure of the ground state is considered to be a canted spin structure with the competition of AFM and FM interactions (shown in Fig. 4 (d)\cite{CrTe_magnetic_structure}). For Cr$_7$Te$_8$, by using different thermal treatment ways, the Cr vacancies can be ordered or disordered with different crystal structures\cite{hashimoto1969magnetic_Cr7Te8_magneticastructure}. For our single crystal, the revealed hexagonal crystal symmetry exhibits the same crystal structure as that of CrTe indicating the disordered arrangements of the vacancies among the successive Cr layers. The saturated magnetic moment is calculated to be 2.4 $\mu_B$ close to that in disordered Cr$_7$Te$_8$ indicating a mixture of FM alignment of Cr$^{2+}$ and AFM alignment of Cr$^{3+}$\cite{hashimoto1969magnetic_Cr7Te8_magneticastructure}.

We first discuss the observed robust THE. This phenomenon is usually attributed to originate from the movement of skyrmions or the non-collinear magnetic structure. In CrTe, according to our former analysis of the ground state, the canted collinear FM order can not break the inversion symmetry. However, the Cr vacancies can host strong magnetic fluctuation leading to non-zero spin chirality locally\cite{Y166_kagome, Tm166_kagome}. With increasing the applied field, magnetic fluctuation is suppressed and spins are aligned along the $c$ axis leading to the THE varnishing. In some reports before\cite{liu2022magnetic_Cr7Te8_skyrmion}, the magnetic moment of Cr undergoes a temperature evolution of tilting from the $c$ axis toward the $ab$ plane and then to the $c$ axis again at the characteristic temperature \textit{T$_s$}, accompanied by the
nonmonotonic change of uniaxial magnetic anisotropy (UMA) constant $\kappa_a$.  The related variations of topological spin textures result in a sign change of THE around \textit{T$_s$} accompanied by the sign change of AHE and OHE. In our measurements,  $\rho_{xy}^{THE-peak}$ exhibits nonlinear temperature dependence suggesting the nonmonotonic evolution of the UMA and dipole interactions. However, in the whole temperature region, the $\rho_{xy}^{THE}$ still keeps the positive values indicating smaller variations of spin chirality originating from magnetic fluctuations in our Cr$_{7-\delta}$Te$_8$.  The robust THE provides large 
values of $\rho_{xy}^{THE-peak}$ ($\sim$ 1 $\mu$$\Omega$$\cdot$$cm$ at 5 K) and high working temperature (room temperature) shown in Fig.3 (c), promising potential for future spintronic applications. 

Now we turn to discuss the AHE and its temperature-dependent evolution of our Cr$_{7-\delta}$Te$_8$. The $R$$_A$ increases slightly as the temperature increases up to 150 K, while above 150 K it decreases promptly as the temperature increases. Eventually, around $T$$_{sr}$=280 K $R$$_A$ reverses the sign. A similar sign change of AHE is also observed in Cr$_3$Te$_4$\cite{Cr3Te4_change_AHE_sign} or Cr$_5$Te$_8$\cite{wang2019magnetic_1Cr5Te8}. In Cr$_{0.87}$Te, it accompanies with the sign of OHE and THE according to former reports\cite{liu2022magnetic_Cr7Te8_skyrmion}. In our Cr$_{7-\delta}$Te$_8$ this sign change of AHE is independent of THE and OHE,  suggesting a transition of magnetic states and interactions which is invisible in DC magnetization measurements. To investigate this transition, temperature-dependent AC  susceptibility $\chi(T) ^ {\prime}$ measurements are performed with various applied fields. As shown in Fig.4 (b) at 0 T, $\chi^ {\prime}(T) $ exhibits a similar temperature dependence with the FM transition as that in $M$($\mu _0H$). With applying the field, the FM transition gradually breaks into two transitions: besides the former one observed around $T$$_{FM}$, a new one emerges at $T$$_{FMN}$ (below $T$$_{FM}$). With increasing the applied field, $T$$_{FMN}$ shifts to low temperatures and eventually stabilizes at $T_{sr}$ with the magnetization saturating. In our DC magnetic measurements, no evidence supports the coexistence and competition of FM and AFM phases in Cr$_{7-\delta}$Te$_8$ like some other Cr-Te system\cite{wang2019magnetic_1Cr5Te8}.  This revealed field-driven transition seems to exhibit more dynamic behaviors instead of the static magnetic character suggesting the emergence of magnetic fluctuations. Around $T$$_{sr}$,
the applied field which can effectively suppress the magnetic fluctuations strongly affects the mutual competition in UMA, AFM and FM interactions.  Although the LR magnetic orders and structure do not seem to change, the transformation of SR and local magnetic states can reconstruct deflections of the carrier motion with the same magnetization direction and drive the sign reversal for AHE \cite{Cr3Te4_change_AHE_sign}.   

Fig. 4(d) shows the phase diagram with a contour map of the Hall resistivity after subtracting the OHE of Cr$_{7-\delta}$Te$_8$. The non-collinear magnetic structure dominates in the low field region, with prominent THE (the yellow region) up to room temperature. The aligned FM structure dominates in the high field region (yellow) with the saturated magnetization. Above $T$$_{sr}$, the AHE reverses the sign suggesting an opposite direction of the carrier motion deflections  (purple). In Cr-Te compounds, the vacancy of Cr is important for the magnetic states and magneto-transport properties. Not only the number of Cr will affect the carrier density and magnetic moment, but also the order of these vacancies can change the crystal structure and bring fluctuations or other SR magnetic orders. In Cr$_{7-\delta}$Te$_8$, the complicated interactions of SR and LR magnetic order would host these rich magnetic states and exotic magneto-transport properties. \cite{Cr5Te6_thin_film}  

\section{IV. SUMMARY}
In summary, we systematically investigate the magnetic states and magneto-transport properties in the centrosymmetric  Cr$_{7-\delta}$Te$_8$. The robust THE persists up to room temperature with a large value attributed to local nonzero spin chirality originating from the LR magnetic orders and fluctuations. The applied field leads to an unreported magnetic transition leading to the sign reversal of AHE. The tunable interactions of LR and SR magnetism in Cr$_{7-\delta}$Te$_8$ can host the rich exotic magnetic states and effects promising potential for future spintronic applications. 

\section{ACKNOWLEDGMENTS}
Project is supported by the National Key Research and Development Program of China (Grant Nos. 2023YFF0718400 and 2023YFA1406500), the National Natural Science Foundation of China (Grant
Nos. U2130101, 92165204), the open research fund of Natural Science Foundation of Guangdong Province (Grant No. 2022A1515010035 and 2021B1515120015), Guangzhou Basic and Applied Basic Research Foundation(Grant No. 202201011798 and 2024A04J6417), Songshan Lake Materials Laboratory (Grant No. 2021SLABFN11), the Open Project of Guangdong Provincial Key Laboratory of Magnetoelectric Physics and Devices (Grant No. 2022B1212010008), the Open Project
of Key Laboratory of Optoelectronic Materials and Technologies (Grant No. OEMT-2023-ZTS-01). The authors thank Dr. Long Jiang from the Instrumental Analysis \& Research Center of Sun Yat-Sen University for single-crystal X-ray diffraction measurements and structural analysis.


\begin{thebibliography}{35}%
\makeatletter
\providecommand \@ifxundefined [1]{%
 \@ifx{#1\undefined}
}%
\providecommand \@ifnum [1]{%
 \ifnum #1\expandafter \@firstoftwo
 \else \expandafter \@secondoftwo
 \fi
}%
\providecommand \@ifx [1]{%
 \ifx #1\expandafter \@firstoftwo
 \else \expandafter \@secondoftwo
 \fi
}%
\providecommand \natexlab [1]{#1}%
\providecommand \enquote  [1]{``#1''}%
\providecommand \bibnamefont  [1]{#1}%
\providecommand \bibfnamefont [1]{#1}%
\providecommand \citenamefont [1]{#1}%
\providecommand \href@noop [0]{\@secondoftwo}%
\providecommand \href [0]{\begingroup \@sanitize@url \@href}%
\providecommand \@href[1]{\@@startlink{#1}\@@href}%
\providecommand \@@href[1]{\endgroup#1\@@endlink}%
\providecommand \@sanitize@url [0]{\catcode `\\12\catcode `\$12\catcode
  `\&12\catcode `\#12\catcode `\^12\catcode `\_12\catcode `\%12\relax}%
\providecommand \@@startlink[1]{}%
\providecommand \@@endlink[0]{}%
\providecommand \url  [0]{\begingroup\@sanitize@url \@url }%
\providecommand \@url [1]{\endgroup\@href {#1}{\urlprefix }}%
\providecommand \urlprefix  [0]{URL }%
\providecommand \Eprint [0]{\href }%
\providecommand \doibase [0]{https://doi.org/}%
\providecommand \selectlanguage [0]{\@gobble}%
\providecommand \bibinfo  [0]{\@secondoftwo}%
\providecommand \bibfield  [0]{\@secondoftwo}%
\providecommand \translation [1]{[#1]}%
\providecommand \BibitemOpen [0]{}%
\providecommand \bibitemStop [0]{}%
\providecommand \bibitemNoStop [0]{.\EOS\space}%
\providecommand \EOS [0]{\spacefactor3000\relax}%
\providecommand \BibitemShut  [1]{\csname bibitem#1\endcsname}%
\let\auto@bib@innerbib\@empty
\bibitem [{\citenamefont {Tokura}\ \emph {et~al.}(2019)\citenamefont {Tokura},
  \citenamefont {Yasuda},\ and\ \citenamefont
  {Tsukazaki}}]{tokura2019magnetic_topological}%
  \BibitemOpen
  \bibfield  {author} {\bibinfo {author} {\bibfnamefont {Y.}~\bibnamefont
  {Tokura}}, \bibinfo {author} {\bibfnamefont {K.}~\bibnamefont {Yasuda}},\
  and\ \bibinfo {author} {\bibfnamefont {A.}~\bibnamefont {Tsukazaki}},\
  }\bibfield  {title} {\bibinfo {title} {Magnetic topological insulators},\
  }\href@noop {} {\bibfield  {journal} {\bibinfo  {journal} {Nat. Rev. Phys.}\
  }\textbf {\bibinfo {volume} {1}},\ \bibinfo {pages} {126} (\bibinfo {year}
  {2019})}\BibitemShut {NoStop}%
\bibitem [{\citenamefont {Narang}\ \emph {et~al.}(2021)\citenamefont {Narang},
  \citenamefont {Garcia},\ and\ \citenamefont
  {Felser}}]{narang2021topology_topological2}%
  \BibitemOpen
  \bibfield  {author} {\bibinfo {author} {\bibfnamefont {P.}~\bibnamefont
  {Narang}}, \bibinfo {author} {\bibfnamefont {C.~A.}\ \bibnamefont {Garcia}},\
  and\ \bibinfo {author} {\bibfnamefont {C.}~\bibnamefont {Felser}},\
  }\bibfield  {title} {\bibinfo {title} {The topology of electronic band
  structures},\ }\href@noop {} {\bibfield  {journal} {\bibinfo  {journal} {Nat.
  Mater.}\ }\textbf {\bibinfo {volume} {20}},\ \bibinfo {pages} {293} (\bibinfo
  {year} {2021})}\BibitemShut {NoStop}%
\bibitem [{\citenamefont {Ndiaye}\ \emph {et~al.}(2017)\citenamefont {Ndiaye},
  \citenamefont {Akosa},\ and\ \citenamefont
  {Manchon}}]{ndiaye2017topological_skyrmion1}%
  \BibitemOpen
  \bibfield  {author} {\bibinfo {author} {\bibfnamefont {P.~B.}\ \bibnamefont
  {Ndiaye}}, \bibinfo {author} {\bibfnamefont {C.~A.}\ \bibnamefont {Akosa}},\
  and\ \bibinfo {author} {\bibfnamefont {A.}~\bibnamefont {Manchon}},\
  }\bibfield  {title} {\bibinfo {title} {Topological hall and spin hall effects
  in disordered skyrmionic textures},\ }\href@noop {} {\bibfield  {journal}
  {\bibinfo  {journal} {Phys. Rev. B}\ }\textbf {\bibinfo {volume} {95}},\
  \bibinfo {pages} {064426} (\bibinfo {year} {2017})}\BibitemShut {NoStop}%
\bibitem [{\citenamefont {Leroux}\ \emph {et~al.}(2018)\citenamefont {Leroux},
  \citenamefont {Stolt}, \citenamefont {Jin}, \citenamefont {Pete},
  \citenamefont {Reichhardt},\ and\ \citenamefont
  {Maiorov}}]{leroux2018skyrmion2}%
  \BibitemOpen
  \bibfield  {author} {\bibinfo {author} {\bibfnamefont {M.}~\bibnamefont
  {Leroux}}, \bibinfo {author} {\bibfnamefont {M.~J.}\ \bibnamefont {Stolt}},
  \bibinfo {author} {\bibfnamefont {S.}~\bibnamefont {Jin}}, \bibinfo {author}
  {\bibfnamefont {D.~V.}\ \bibnamefont {Pete}}, \bibinfo {author}
  {\bibfnamefont {C.}~\bibnamefont {Reichhardt}},\ and\ \bibinfo {author}
  {\bibfnamefont {B.}~\bibnamefont {Maiorov}},\ }\bibfield  {title} {\bibinfo
  {title} {Skyrmion lattice topological hall effect near room temperature},\
  }\href@noop {} {\bibfield  {journal} {\bibinfo  {journal} {Sci. Rep.}\
  }\textbf {\bibinfo {volume} {8}},\ \bibinfo {pages} {15510} (\bibinfo {year}
  {2018})}\BibitemShut {NoStop}%
\bibitem [{\citenamefont {Seshadri}\ and\ \citenamefont
  {Sen}(2018)}]{seshadri2018topological_DMI1}%
  \BibitemOpen
  \bibfield  {author} {\bibinfo {author} {\bibfnamefont {R.}~\bibnamefont
  {Seshadri}}\ and\ \bibinfo {author} {\bibfnamefont {D.}~\bibnamefont {Sen}},\
  }\bibfield  {title} {\bibinfo {title} {{Topological magnons in a
  kagome-lattice spin system with X X Z and Dzyaloshinskii-Moriya
  interactions}},\ }\href@noop {} {\bibfield  {journal} {\bibinfo  {journal}
  {Phys. Rev. B}\ }\textbf {\bibinfo {volume} {97}},\ \bibinfo {pages} {134411}
  (\bibinfo {year} {2018})}\BibitemShut {NoStop}%
\bibitem [{\citenamefont {Fern{\'a}ndez-Pacheco}\ \emph
  {et~al.}(2019)\citenamefont {Fern{\'a}ndez-Pacheco}, \citenamefont
  {Vedmedenko}, \citenamefont {Ummelen}, \citenamefont {Mansell}, \citenamefont
  {Petit},\ and\ \citenamefont {Cowburn}}]{fernandez2019symmetry_DMI2}%
  \BibitemOpen
  \bibfield  {author} {\bibinfo {author} {\bibfnamefont {A.}~\bibnamefont
  {Fern{\'a}ndez-Pacheco}}, \bibinfo {author} {\bibfnamefont {E.}~\bibnamefont
  {Vedmedenko}}, \bibinfo {author} {\bibfnamefont {F.}~\bibnamefont {Ummelen}},
  \bibinfo {author} {\bibfnamefont {R.}~\bibnamefont {Mansell}}, \bibinfo
  {author} {\bibfnamefont {D.}~\bibnamefont {Petit}},\ and\ \bibinfo {author}
  {\bibfnamefont {R.~P.}\ \bibnamefont {Cowburn}},\ }\bibfield  {title}
  {\bibinfo {title} {{Symmetry-breaking interlayer Dzyaloshinskii--Moriya
  interactions in synthetic antiferromagnets}},\ }\href@noop {} {\bibfield
  {journal} {\bibinfo  {journal} {Nat. Mater.}\ }\textbf {\bibinfo {volume}
  {18}},\ \bibinfo {pages} {679} (\bibinfo {year} {2019})}\BibitemShut
  {NoStop}%
\bibitem [{\citenamefont {Zhang}\ \emph {et~al.}(2018)\citenamefont {Zhang},
  \citenamefont {Lu}, \citenamefont {Zhu}, \citenamefont {Tan}, \citenamefont
  {Feng}, \citenamefont {Liu}, \citenamefont {Zhang}, \citenamefont {Chen},
  \citenamefont {Liu}, \citenamefont {Luo} \emph
  {et~al.}}]{zhang2018emergence_Fe3GeTe2}%
  \BibitemOpen
  \bibfield  {author} {\bibinfo {author} {\bibfnamefont {Y.}~\bibnamefont
  {Zhang}}, \bibinfo {author} {\bibfnamefont {H.}~\bibnamefont {Lu}}, \bibinfo
  {author} {\bibfnamefont {X.}~\bibnamefont {Zhu}}, \bibinfo {author}
  {\bibfnamefont {S.}~\bibnamefont {Tan}}, \bibinfo {author} {\bibfnamefont
  {W.}~\bibnamefont {Feng}}, \bibinfo {author} {\bibfnamefont {Q.}~\bibnamefont
  {Liu}}, \bibinfo {author} {\bibfnamefont {W.}~\bibnamefont {Zhang}}, \bibinfo
  {author} {\bibfnamefont {Q.}~\bibnamefont {Chen}}, \bibinfo {author}
  {\bibfnamefont {Y.}~\bibnamefont {Liu}}, \bibinfo {author} {\bibfnamefont
  {X.}~\bibnamefont {Luo}}, \emph {et~al.},\ }\bibfield  {title} {\bibinfo
  {title} {{Emergence of Kondo lattice behavior in a van der Waals itinerant
  ferromagnet, $\mathrm{Fe}_{3}\mathrm{Ge}\mathrm{Te}_{2}$}},\ }\href@noop
  {} {\bibfield  {journal} {\bibinfo  {journal} {Sci. Adv.}\ }\textbf {\bibinfo
  {volume} {4}},\ \bibinfo {pages} {eaao6791} (\bibinfo {year}
  {2018})}\BibitemShut {NoStop}%
\bibitem [{\citenamefont {Carteaux}\ \emph {et~al.}(1995)\citenamefont
  {Carteaux}, \citenamefont {Brunet}, \citenamefont {Ouvrard},\ and\
  \citenamefont {Andre}}]{carteaux1995crystallographic_1Cr2Ge2Te6}%
  \BibitemOpen
  \bibfield  {author} {\bibinfo {author} {\bibfnamefont {V.}~\bibnamefont
  {Carteaux}}, \bibinfo {author} {\bibfnamefont {D.}~\bibnamefont {Brunet}},
  \bibinfo {author} {\bibfnamefont {G.}~\bibnamefont {Ouvrard}},\ and\ \bibinfo
  {author} {\bibfnamefont {G.}~\bibnamefont {Andre}},\ }\bibfield  {title}
  {\bibinfo {title} {{Crystallographic, magnetic and electronic structures of a
  new layered ferromagnetic compound
  $\mathrm{Cr}_{2}\mathrm{Ge}_{2}\mathrm{Te}_{6}$}},\ }\href@noop {}
  {\bibfield  {journal} {\bibinfo  {journal} {J. Phys. Condens. Matter.}\
  }\textbf {\bibinfo {volume} {7}},\ \bibinfo {pages} {69} (\bibinfo {year}
  {1995})}\BibitemShut {NoStop}%
\bibitem [{\citenamefont {Zhang}\ \emph {et~al.}(2015)\citenamefont {Zhang},
  \citenamefont {Zhao}, \citenamefont {Yao},\ and\ \citenamefont
  {Yang}}]{zhang2015robust_2Cr2Ge2Te6}%
  \BibitemOpen
  \bibfield  {author} {\bibinfo {author} {\bibfnamefont {J.}~\bibnamefont
  {Zhang}}, \bibinfo {author} {\bibfnamefont {B.}~\bibnamefont {Zhao}},
  \bibinfo {author} {\bibfnamefont {Y.}~\bibnamefont {Yao}},\ and\ \bibinfo
  {author} {\bibfnamefont {Z.}~\bibnamefont {Yang}},\ }\bibfield  {title}
  {\bibinfo {title} {{Robust quantum anomalous Hall effect in graphene-based
  van der Waals heterostructures}},\ }\href@noop {} {\bibfield  {journal}
  {\bibinfo  {journal} {Phys. Rev. B}\ }\textbf {\bibinfo {volume} {92}},\
  \bibinfo {pages} {165418} (\bibinfo {year} {2015})}\BibitemShut {NoStop}%
\bibitem [{\citenamefont {Zhang}\ \emph {et~al.}(2019)\citenamefont {Zhang},
  \citenamefont {Shi}, \citenamefont {Zhu}, \citenamefont {Xing}, \citenamefont
  {Zhang},\ and\ \citenamefont {Wang}}]{zhang2019topological_1MnBi2Te4}%
  \BibitemOpen
  \bibfield  {author} {\bibinfo {author} {\bibfnamefont {D.}~\bibnamefont
  {Zhang}}, \bibinfo {author} {\bibfnamefont {M.}~\bibnamefont {Shi}}, \bibinfo
  {author} {\bibfnamefont {T.}~\bibnamefont {Zhu}}, \bibinfo {author}
  {\bibfnamefont {D.}~\bibnamefont {Xing}}, \bibinfo {author} {\bibfnamefont
  {H.}~\bibnamefont {Zhang}},\ and\ \bibinfo {author} {\bibfnamefont
  {J.}~\bibnamefont {Wang}},\ }\bibfield  {title} {\bibinfo {title}
  {{Topological axion states in the magnetic insulator
  $\mathrm{Mn}\mathrm{Bi}_2\mathrm{Te}_{4}$ with the quantized
  magnetoelectric effect}},\ }\href@noop {} {\bibfield  {journal} {\bibinfo
  {journal} {Phys. Rev. Lett.}\ }\textbf {\bibinfo {volume} {122}},\ \bibinfo
  {pages} {206401} (\bibinfo {year} {2019})}\BibitemShut {NoStop}%
\bibitem [{\citenamefont {Deng}\ \emph {et~al.}(2020)\citenamefont {Deng},
  \citenamefont {Yu}, \citenamefont {Shi}, \citenamefont {Guo}, \citenamefont
  {Xu}, \citenamefont {Wang}, \citenamefont {Chen},\ and\ \citenamefont
  {Zhang}}]{deng2020quantum_2MnBi2Te4}%
  \BibitemOpen
  \bibfield  {author} {\bibinfo {author} {\bibfnamefont {Y.}~\bibnamefont
  {Deng}}, \bibinfo {author} {\bibfnamefont {Y.}~\bibnamefont {Yu}}, \bibinfo
  {author} {\bibfnamefont {M.~Z.}\ \bibnamefont {Shi}}, \bibinfo {author}
  {\bibfnamefont {Z.}~\bibnamefont {Guo}}, \bibinfo {author} {\bibfnamefont
  {Z.}~\bibnamefont {Xu}}, \bibinfo {author} {\bibfnamefont {J.}~\bibnamefont
  {Wang}}, \bibinfo {author} {\bibfnamefont {X.~H.}\ \bibnamefont {Chen}},\
  and\ \bibinfo {author} {\bibfnamefont {Y.}~\bibnamefont {Zhang}},\ }\bibfield
   {title} {\bibinfo {title} {{Quantum anomalous Hall effect in intrinsic
  magnetic topological insulator
  $\mathrm{Mn}\mathrm{Bi}_2\mathrm{Te}_{4}$}},\ }\href@noop {} {\bibfield
  {journal} {\bibinfo  {journal} {Science}\ }\textbf {\bibinfo {volume}
  {367}},\ \bibinfo {pages} {895} (\bibinfo {year} {2020})}\BibitemShut
  {NoStop}%
\bibitem [{\citenamefont {Yi}\ \emph {et~al.}(2023)\citenamefont {Yi},
  \citenamefont {Zheng}, \citenamefont {Pan}, \citenamefont {Zhang},
  \citenamefont {Wang}, \citenamefont {Chen}, \citenamefont {Wu}, \citenamefont
  {Liang}, \citenamefont {Mei}, \citenamefont {Wu}, \citenamefont {Yang},
  \citenamefont {Cheng}, \citenamefont {Wang},\ and\ \citenamefont
  {Shen}}]{EuZn2As2_short_range_order}%
  \BibitemOpen
  \bibfield  {author} {\bibinfo {author} {\bibfnamefont {E.}~\bibnamefont
  {Yi}}, \bibinfo {author} {\bibfnamefont {D.~F.}\ \bibnamefont {Zheng}},
  \bibinfo {author} {\bibfnamefont {F.}~\bibnamefont {Pan}}, \bibinfo {author}
  {\bibfnamefont {H.}~\bibnamefont {Zhang}}, \bibinfo {author} {\bibfnamefont
  {B.}~\bibnamefont {Wang}}, \bibinfo {author} {\bibfnamefont {B.}~\bibnamefont
  {Chen}}, \bibinfo {author} {\bibfnamefont {D.}~\bibnamefont {Wu}}, \bibinfo
  {author} {\bibfnamefont {H.}~\bibnamefont {Liang}}, \bibinfo {author}
  {\bibfnamefont {Z.~X.}\ \bibnamefont {Mei}}, \bibinfo {author} {\bibfnamefont
  {H.}~\bibnamefont {Wu}}, \bibinfo {author} {\bibfnamefont {S.~A.}\
  \bibnamefont {Yang}}, \bibinfo {author} {\bibfnamefont {P.}~\bibnamefont
  {Cheng}}, \bibinfo {author} {\bibfnamefont {M.}~\bibnamefont {Wang}},\ and\
  \bibinfo {author} {\bibfnamefont {B.}~\bibnamefont {Shen}},\ }\bibfield
  {title} {\bibinfo {title} {{Topological Hall effect driven by short-range
  magnetic order in ${\mathrm{EuZn}}_{2}{\mathrm{As}}_{2}$}},\ }\href
  {https://doi.org/10.1103/PhysRevB.107.035142} {\bibfield  {journal} {\bibinfo
   {journal} {Phys. Rev. B}\ }\textbf {\bibinfo {volume} {107}},\ \bibinfo
  {pages} {035142} (\bibinfo {year} {2023})}\BibitemShut {NoStop}%
\bibitem [{\citenamefont {Ma}\ \emph {et~al.}(2019)\citenamefont {Ma},
  \citenamefont {Nie}, \citenamefont {Yi}, \citenamefont {Jandke},
  \citenamefont {Shang}, \citenamefont {Yao}, \citenamefont {Naamneh},
  \citenamefont {Yan}, \citenamefont {Sun}, \citenamefont {Chikina},
  \citenamefont {Strocov}, \citenamefont {Medarde}, \citenamefont {Song},
  \citenamefont {Xiong}, \citenamefont {Xu}, \citenamefont {Wulfhekel},
  \citenamefont {Mesot}, \citenamefont {Reticcioli}, \citenamefont {Franchini},
  \citenamefont {Mudry}, \citenamefont {Müller}, \citenamefont {Shi},
  \citenamefont {Qian}, \citenamefont {Ding},\ and\ \citenamefont
  {Shi}}]{EuCd2As2_spin_fluctuation}%
  \BibitemOpen
  \bibfield  {author} {\bibinfo {author} {\bibfnamefont {J.-Z.}\ \bibnamefont
  {Ma}}, \bibinfo {author} {\bibfnamefont {S.~M.}\ \bibnamefont {Nie}},
  \bibinfo {author} {\bibfnamefont {C.~J.}\ \bibnamefont {Yi}}, \bibinfo
  {author} {\bibfnamefont {J.}~\bibnamefont {Jandke}}, \bibinfo {author}
  {\bibfnamefont {T.}~\bibnamefont {Shang}}, \bibinfo {author} {\bibfnamefont
  {M.~Y.}\ \bibnamefont {Yao}}, \bibinfo {author} {\bibfnamefont
  {M.}~\bibnamefont {Naamneh}}, \bibinfo {author} {\bibfnamefont {L.~Q.}\
  \bibnamefont {Yan}}, \bibinfo {author} {\bibfnamefont {Y.}~\bibnamefont
  {Sun}}, \bibinfo {author} {\bibfnamefont {A.}~\bibnamefont {Chikina}},
  \bibinfo {author} {\bibfnamefont {V.~N.}\ \bibnamefont {Strocov}}, \bibinfo
  {author} {\bibfnamefont {M.}~\bibnamefont {Medarde}}, \bibinfo {author}
  {\bibfnamefont {M.}~\bibnamefont {Song}}, \bibinfo {author} {\bibfnamefont
  {Y.-M.}\ \bibnamefont {Xiong}}, \bibinfo {author} {\bibfnamefont
  {G.}~\bibnamefont {Xu}}, \bibinfo {author} {\bibfnamefont {W.}~\bibnamefont
  {Wulfhekel}}, \bibinfo {author} {\bibfnamefont {J.}~\bibnamefont {Mesot}},
  \bibinfo {author} {\bibfnamefont {M.}~\bibnamefont {Reticcioli}}, \bibinfo
  {author} {\bibfnamefont {C.}~\bibnamefont {Franchini}}, \bibinfo {author}
  {\bibfnamefont {C.}~\bibnamefont {Mudry}}, \bibinfo {author} {\bibfnamefont
  {M.}~\bibnamefont {Müller}}, \bibinfo {author} {\bibfnamefont {Y.~G.}\
  \bibnamefont {Shi}}, \bibinfo {author} {\bibfnamefont {T.}~\bibnamefont
  {Qian}}, \bibinfo {author} {\bibfnamefont {H.}~\bibnamefont {Ding}},\ and\
  \bibinfo {author} {\bibfnamefont {M.}~\bibnamefont {Shi}},\ }\bibfield
  {title} {\bibinfo {title} {{Spin fluctuation induced Weyl semimetal state in
  the paramagnetic phase of $\mathrm{Eu}\mathrm{Cd}_2\mathrm{As}_2$}},\
  }\href {https://doi.org/10.1126/sciadv.aaw4718} {\bibfield  {journal}
  {\bibinfo  {journal} {Sci. Adv.}\ }\textbf {\bibinfo {volume} {5}},\ \bibinfo
  {pages} {eaaw4718} (\bibinfo {year} {2019})}\BibitemShut {NoStop}%
\bibitem [{\citenamefont {Wang}\ \emph {et~al.}(2022)\citenamefont {Wang},
  \citenamefont {Yi}, \citenamefont {Li}, \citenamefont {Qin}, \citenamefont
  {Hu}, \citenamefont {Shen},\ and\ \citenamefont {Wang}}]{Tm166_kagome}%
  \BibitemOpen
  \bibfield  {author} {\bibinfo {author} {\bibfnamefont {B.}~\bibnamefont
  {Wang}}, \bibinfo {author} {\bibfnamefont {E.}~\bibnamefont {Yi}}, \bibinfo
  {author} {\bibfnamefont {L.}~\bibnamefont {Li}}, \bibinfo {author}
  {\bibfnamefont {J.}~\bibnamefont {Qin}}, \bibinfo {author} {\bibfnamefont
  {B.-F.}\ \bibnamefont {Hu}}, \bibinfo {author} {\bibfnamefont
  {B.}~\bibnamefont {Shen}},\ and\ \bibinfo {author} {\bibfnamefont
  {M.}~\bibnamefont {Wang}},\ }\bibfield  {title} {\bibinfo {title}
  {{Magnetotransport properties of the kagome magnet
  ${\mathrm{TmMn}}_{6}{\mathrm{Sn}}_{6}$}},\ }\href
  {https://doi.org/10.1103/PhysRevB.106.125107} {\bibfield  {journal} {\bibinfo
   {journal} {Phys. Rev. B}\ }\textbf {\bibinfo {volume} {106}},\ \bibinfo
  {pages} {125107} (\bibinfo {year} {2022})}\BibitemShut {NoStop}%
\bibitem [{\citenamefont {Sun}\ \emph {et~al.}(2021)\citenamefont {Sun},
  \citenamefont {Chen}, \citenamefont {Hou}, \citenamefont {Wang},
  \citenamefont {Gong}, \citenamefont {Huo}, \citenamefont {Li}, \citenamefont
  {Yu}, \citenamefont {Cai}, \citenamefont {Liu}, \citenamefont {Wu},
  \citenamefont {Yao},\ and\ \citenamefont {Wang}}]{EuSn2As2_variation}%
  \BibitemOpen
  \bibfield  {author} {\bibinfo {author} {\bibfnamefont {H.}~\bibnamefont
  {Sun}}, \bibinfo {author} {\bibfnamefont {C.}~\bibnamefont {Chen}}, \bibinfo
  {author} {\bibfnamefont {Y.}~\bibnamefont {Hou}}, \bibinfo {author}
  {\bibfnamefont {W.}~\bibnamefont {Wang}}, \bibinfo {author} {\bibfnamefont
  {Y.}~\bibnamefont {Gong}}, \bibinfo {author} {\bibfnamefont {M.}~\bibnamefont
  {Huo}}, \bibinfo {author} {\bibfnamefont {L.}~\bibnamefont {Li}}, \bibinfo
  {author} {\bibfnamefont {J.}~\bibnamefont {Yu}}, \bibinfo {author}
  {\bibfnamefont {W.}~\bibnamefont {Cai}}, \bibinfo {author} {\bibfnamefont
  {N.}~\bibnamefont {Liu}}, \bibinfo {author} {\bibfnamefont {R.}~\bibnamefont
  {Wu}}, \bibinfo {author} {\bibfnamefont {D.-X.}\ \bibnamefont {Yao}},\ and\
  \bibinfo {author} {\bibfnamefont {M.}~\bibnamefont {Wang}},\ }\bibfield
  {title} {\bibinfo {title} {{Magnetism variation of the compressed
  antiferromagnetic topological insulator EuSn$_2$As$_2$}},\ }\href
  {https://doi.org/10.1007/s11433-021-1760-x} {\bibfield  {journal} {\bibinfo
  {journal} {SCI CHINA PHYS MECH}\ }\textbf {\bibinfo {volume} {64}},\ \bibinfo
  {pages} {118211} (\bibinfo {year} {2021})}\BibitemShut {NoStop}%
\bibitem [{\citenamefont {Nisoli}\ \emph {et~al.}(2013)\citenamefont {Nisoli},
  \citenamefont {Moessner},\ and\ \citenamefont
  {Schiffer}}]{nisoli2013colloquium_SRandLR1}%
  \BibitemOpen
  \bibfield  {author} {\bibinfo {author} {\bibfnamefont {C.}~\bibnamefont
  {Nisoli}}, \bibinfo {author} {\bibfnamefont {R.}~\bibnamefont {Moessner}},\
  and\ \bibinfo {author} {\bibfnamefont {P.}~\bibnamefont {Schiffer}},\
  }\bibfield  {title} {\bibinfo {title} {{Colloquium: Artificial spin ice:
  Designing and imaging magnetic frustration}},\ }\href@noop {} {\bibfield
  {journal} {\bibinfo  {journal} {Rev. Mod. Phys.}\ }\textbf {\bibinfo {volume}
  {85}},\ \bibinfo {pages} {1473} (\bibinfo {year} {2013})}\BibitemShut
  {NoStop}%
\bibitem [{\citenamefont {Mydosh}(2015)}]{mydosh2015spin_SRandLR2}%
  \BibitemOpen
  \bibfield  {author} {\bibinfo {author} {\bibfnamefont {J.}~\bibnamefont
  {Mydosh}},\ }\bibfield  {title} {\bibinfo {title} {{Spin glasses: redux: an
  updated experimental/materials survey}},\ }\href@noop {} {\bibfield
  {journal} {\bibinfo  {journal} {Rep. Prog. Phys.}\ }\textbf {\bibinfo
  {volume} {78}},\ \bibinfo {pages} {052501} (\bibinfo {year}
  {2015})}\BibitemShut {NoStop}%
\bibitem [{\citenamefont {Ipser}\ \emph {et~al.}(1983)\citenamefont {Ipser},
  \citenamefont {Komarek},\ and\ \citenamefont {Klepp}}]{CrTe_phase_structure}%
  \BibitemOpen
  \bibfield  {author} {\bibinfo {author} {\bibfnamefont {H.}~\bibnamefont
  {Ipser}}, \bibinfo {author} {\bibfnamefont {K.~L.}\ \bibnamefont {Komarek}},\
  and\ \bibinfo {author} {\bibfnamefont {K.~O.}\ \bibnamefont {Klepp}},\
  }\bibfield  {title} {\bibinfo {title} {{Transition metal-chalcogen systems
  viii: The $\mathrm{Cr}-\mathrm{Te}$ phase diagram}},\ }\href
  {https://doi.org/https://doi.org/10.1016/0022-5088(83)90493-9} {\bibfield
  {journal} {\bibinfo  {journal} {J. less-common met.}\ }\textbf {\bibinfo
  {volume} {92}},\ \bibinfo {pages} {265} (\bibinfo {year} {1983})}\BibitemShut
  {NoStop}%
\bibitem [{\citenamefont {Hashimoto}\ and\ \citenamefont
  {Yamaguchi}(1969)}]{hashimoto1969magnetic_Cr7Te8_magneticastructure}%
  \BibitemOpen
  \bibfield  {author} {\bibinfo {author} {\bibfnamefont {T.}~\bibnamefont
  {Hashimoto}}\ and\ \bibinfo {author} {\bibfnamefont {M.}~\bibnamefont
  {Yamaguchi}},\ }\bibfield  {title} {\bibinfo {title} {{Magnetic Properties of
  $\mathrm{Cr}_7\mathrm{Te}_{8}$}},\ }\href@noop {} {\bibfield  {journal}
  {\bibinfo  {journal} {J. Phys. Soc. Japan}\ }\textbf {\bibinfo {volume}
  {27}},\ \bibinfo {pages} {1121} (\bibinfo {year} {1969})}\BibitemShut
  {NoStop}%
\bibitem [{\citenamefont {Akram}\ and\ \citenamefont
  {Nazar}(1983)}]{akram1983magnetic_tunnable_structure}%
  \BibitemOpen
  \bibfield  {author} {\bibinfo {author} {\bibfnamefont {M.}~\bibnamefont
  {Akram}}\ and\ \bibinfo {author} {\bibfnamefont {F.~M.}\ \bibnamefont
  {Nazar}},\ }\bibfield  {title} {\bibinfo {title} {{Magnetic properties of
  $\mathrm{Cr}\mathrm{Te}$, $\mathrm{Cr}_{23}\mathrm{Te}_{24}$,
  $\mathrm{Cr}_7\mathrm{Te}_{8}$, $\mathrm{Cr}_5\mathrm{Te}_{6}$, and
  $\mathrm{Cr}_3\mathrm{Te}_{4}$ compounds}},\ }\href@noop {} {\bibfield
  {journal} {\bibinfo  {journal} {J. Mater. Sci.}\ }\textbf {\bibinfo {volume}
  {18}},\ \bibinfo {pages} {423} (\bibinfo {year} {1983})}\BibitemShut
  {NoStop}%
\bibitem [{\citenamefont {Dijkstra}\ \emph {et~al.}(1989)\citenamefont
  {Dijkstra}, \citenamefont {Weitering}, \citenamefont {Van~Bruggen},
  \citenamefont {Haas},\ and\ \citenamefont
  {De~Groot}}]{dijkstra1989band_calculation}%
  \BibitemOpen
  \bibfield  {author} {\bibinfo {author} {\bibfnamefont {J.}~\bibnamefont
  {Dijkstra}}, \bibinfo {author} {\bibfnamefont {H.}~\bibnamefont {Weitering}},
  \bibinfo {author} {\bibfnamefont {C.}~\bibnamefont {Van~Bruggen}}, \bibinfo
  {author} {\bibfnamefont {C.}~\bibnamefont {Haas}},\ and\ \bibinfo {author}
  {\bibfnamefont {R.}~\bibnamefont {De~Groot}},\ }\bibfield  {title} {\bibinfo
  {title} {{Band-structure calculations, and magnetic and transport properties
  of ferromagnetic chromium tellurides ($\mathrm{Cr}\mathrm{Te}$,
  $\mathrm{Cr}_3\mathrm{Te}_{4}$, $\mathrm{Cr}_2\mathrm{Te}_{3}$)}},\
  }\href@noop {} {\bibfield  {journal} {\bibinfo  {journal} {J. Phys. Condens.
  Matter.}\ }\textbf {\bibinfo {volume} {1}},\ \bibinfo {pages} {9141}
  (\bibinfo {year} {1989})}\BibitemShut {NoStop}%
\bibitem [{\citenamefont {Wang}\ \emph {et~al.}(2019)\citenamefont {Wang},
  \citenamefont {Yan}, \citenamefont {Li}, \citenamefont {Wang}, \citenamefont
  {Song}, \citenamefont {Song}, \citenamefont {Li}, \citenamefont {Chen},
  \citenamefont {Qin}, \citenamefont {Ling} \emph
  {et~al.}}]{wang2019magnetic_1Cr5Te8}%
  \BibitemOpen
  \bibfield  {author} {\bibinfo {author} {\bibfnamefont {Y.}~\bibnamefont
  {Wang}}, \bibinfo {author} {\bibfnamefont {J.}~\bibnamefont {Yan}}, \bibinfo
  {author} {\bibfnamefont {J.}~\bibnamefont {Li}}, \bibinfo {author}
  {\bibfnamefont {S.}~\bibnamefont {Wang}}, \bibinfo {author} {\bibfnamefont
  {M.}~\bibnamefont {Song}}, \bibinfo {author} {\bibfnamefont {J.}~\bibnamefont
  {Song}}, \bibinfo {author} {\bibfnamefont {Z.}~\bibnamefont {Li}}, \bibinfo
  {author} {\bibfnamefont {K.}~\bibnamefont {Chen}}, \bibinfo {author}
  {\bibfnamefont {Y.}~\bibnamefont {Qin}}, \bibinfo {author} {\bibfnamefont
  {L.}~\bibnamefont {Ling}}, \emph {et~al.},\ }\bibfield  {title} {\bibinfo
  {title} {{Magnetic anisotropy and topological Hall effect in the trigonal
  chromium tellurides $\mathrm{Cr}_5\mathrm{Te}_{8}$}},\ }\href@noop {}
  {\bibfield  {journal} {\bibinfo  {journal} {Phys. Rev. B}\ }\textbf {\bibinfo
  {volume} {100}},\ \bibinfo {pages} {024434} (\bibinfo {year}
  {2019})}\BibitemShut {NoStop}%
\bibitem [{\citenamefont {Jiang}\ \emph {et~al.}(2020)\citenamefont {Jiang},
  \citenamefont {Luo}, \citenamefont {Yan}, \citenamefont {Gao}, \citenamefont
  {Wang}, \citenamefont {Zhao}, \citenamefont {Sun}, \citenamefont {Si},
  \citenamefont {Lu}, \citenamefont {Tong}, \citenamefont {Zhu}, \citenamefont
  {Song},\ and\ \citenamefont {Sun}}]{Magnetic_anisotropy_AHE_MCr5Te8}%
  \BibitemOpen
  \bibfield  {author} {\bibinfo {author} {\bibfnamefont {Z.~Z.}\ \bibnamefont
  {Jiang}}, \bibinfo {author} {\bibfnamefont {X.}~\bibnamefont {Luo}}, \bibinfo
  {author} {\bibfnamefont {J.}~\bibnamefont {Yan}}, \bibinfo {author}
  {\bibfnamefont {J.~J.}\ \bibnamefont {Gao}}, \bibinfo {author} {\bibfnamefont
  {W.}~\bibnamefont {Wang}}, \bibinfo {author} {\bibfnamefont {G.~C.}\
  \bibnamefont {Zhao}}, \bibinfo {author} {\bibfnamefont {Y.}~\bibnamefont
  {Sun}}, \bibinfo {author} {\bibfnamefont {J.~G.}\ \bibnamefont {Si}},
  \bibinfo {author} {\bibfnamefont {W.~J.}\ \bibnamefont {Lu}}, \bibinfo
  {author} {\bibfnamefont {P.}~\bibnamefont {Tong}}, \bibinfo {author}
  {\bibfnamefont {X.~B.}\ \bibnamefont {Zhu}}, \bibinfo {author} {\bibfnamefont
  {W.~H.}\ \bibnamefont {Song}},\ and\ \bibinfo {author} {\bibfnamefont
  {Y.~P.}\ \bibnamefont {Sun}},\ }\bibfield  {title} {\bibinfo {title}
  {{Magnetic anisotropy and anomalous Hall effect in monoclinic single crystal
  $\mathrm{Cr}_{5}\mathrm{Te}_{8}$}},\ }\href
  {https://doi.org/10.1103/PhysRevB.102.144433} {\bibfield  {journal} {\bibinfo
   {journal} {Phys. Rev. B}\ }\textbf {\bibinfo {volume} {102}},\ \bibinfo
  {pages} {144433} (\bibinfo {year} {2020})}\BibitemShut {NoStop}%
\bibitem [{\citenamefont {Zhang}\ \emph {et~al.}(2021)\citenamefont {Zhang},
  \citenamefont {Lu}, \citenamefont {Liu}, \citenamefont {Niu}, \citenamefont
  {Sun}, \citenamefont {Cook}, \citenamefont {Vaninger}, \citenamefont
  {Miceli}, \citenamefont {Singh}, \citenamefont {Lian} \emph
  {et~al.}}]{zhang2021room_CrTe2}%
  \BibitemOpen
  \bibfield  {author} {\bibinfo {author} {\bibfnamefont {X.}~\bibnamefont
  {Zhang}}, \bibinfo {author} {\bibfnamefont {Q.}~\bibnamefont {Lu}}, \bibinfo
  {author} {\bibfnamefont {W.}~\bibnamefont {Liu}}, \bibinfo {author}
  {\bibfnamefont {W.}~\bibnamefont {Niu}}, \bibinfo {author} {\bibfnamefont
  {J.}~\bibnamefont {Sun}}, \bibinfo {author} {\bibfnamefont {J.}~\bibnamefont
  {Cook}}, \bibinfo {author} {\bibfnamefont {M.}~\bibnamefont {Vaninger}},
  \bibinfo {author} {\bibfnamefont {P.~F.}\ \bibnamefont {Miceli}}, \bibinfo
  {author} {\bibfnamefont {D.~J.}\ \bibnamefont {Singh}}, \bibinfo {author}
  {\bibfnamefont {S.-W.}\ \bibnamefont {Lian}}, \emph {et~al.},\ }\bibfield
  {title} {\bibinfo {title} {{Room-temperature intrinsic ferromagnetism in
  epitaxial $\mathrm{Cr}\mathrm{Te}$$_{2}$ ultrathin films}},\ }\href@noop {}
  {\bibfield  {journal} {\bibinfo  {journal} {Nat. Commun.}\ }\textbf {\bibinfo
  {volume} {12}},\ \bibinfo {pages} {2492} (\bibinfo {year}
  {2021})}\BibitemShut {NoStop}%
\bibitem [{\citenamefont {Tang}\ \emph {et~al.}(2022)\citenamefont {Tang},
  \citenamefont {Wang}, \citenamefont {Han}, \citenamefont {Xu}, \citenamefont
  {Zhang}, \citenamefont {Zhu}, \citenamefont {Cao}, \citenamefont {Yang},
  \citenamefont {Fu}, \citenamefont {Yang} \emph
  {et~al.}}]{tang2022phase_2Cr5Te8}%
  \BibitemOpen
  \bibfield  {author} {\bibinfo {author} {\bibfnamefont {B.}~\bibnamefont
  {Tang}}, \bibinfo {author} {\bibfnamefont {X.}~\bibnamefont {Wang}}, \bibinfo
  {author} {\bibfnamefont {M.}~\bibnamefont {Han}}, \bibinfo {author}
  {\bibfnamefont {X.}~\bibnamefont {Xu}}, \bibinfo {author} {\bibfnamefont
  {Z.}~\bibnamefont {Zhang}}, \bibinfo {author} {\bibfnamefont
  {C.}~\bibnamefont {Zhu}}, \bibinfo {author} {\bibfnamefont {X.}~\bibnamefont
  {Cao}}, \bibinfo {author} {\bibfnamefont {Y.}~\bibnamefont {Yang}}, \bibinfo
  {author} {\bibfnamefont {Q.}~\bibnamefont {Fu}}, \bibinfo {author}
  {\bibfnamefont {J.}~\bibnamefont {Yang}}, \emph {et~al.},\ }\bibfield
  {title} {\bibinfo {title} {{Phase engineering of
 $ \mathrm{Cr}_5\mathrm{Te}_{8}$ with colossal anomalous Hall effect}},\
  }\href@noop {} {\bibfield  {journal} {\bibinfo  {journal} {Nat. Electron.}\
  }\textbf {\bibinfo {volume} {5}},\ \bibinfo {pages} {224} (\bibinfo {year}
  {2022})}\BibitemShut {NoStop}%
\bibitem [{\citenamefont {Liu}\ \emph {et~al.}(2022)\citenamefont {Liu},
  \citenamefont {Ding}, \citenamefont {Liang}, \citenamefont {Li},
  \citenamefont {Yao},\ and\ \citenamefont
  {Wang}}]{liu2022magnetic_Cr7Te8_skyrmion}%
  \BibitemOpen
  \bibfield  {author} {\bibinfo {author} {\bibfnamefont {J.}~\bibnamefont
  {Liu}}, \bibinfo {author} {\bibfnamefont {B.}~\bibnamefont {Ding}}, \bibinfo
  {author} {\bibfnamefont {J.}~\bibnamefont {Liang}}, \bibinfo {author}
  {\bibfnamefont {X.}~\bibnamefont {Li}}, \bibinfo {author} {\bibfnamefont
  {Y.}~\bibnamefont {Yao}},\ and\ \bibinfo {author} {\bibfnamefont
  {W.}~\bibnamefont {Wang}},\ }\bibfield  {title} {\bibinfo {title} {{Magnetic
  Skyrmionic bubbles at room temperature and sign reversal of the topological
  Hall effect in a layered ferromagnet $\mathrm{Cr}_{0.87}\mathrm{Te}$}},\
  }\href@noop {} {\bibfield  {journal} {\bibinfo  {journal} {ACS nano}\
  }\textbf {\bibinfo {volume} {16}},\ \bibinfo {pages} {13911} (\bibinfo {year}
  {2022})}\BibitemShut {NoStop}%
\bibitem [{\citenamefont {Zhang}\ \emph {et~al.}(2023)\citenamefont {Zhang},
  \citenamefont {Liu}, \citenamefont {Zhang}, \citenamefont {Yuan},
  \citenamefont {Wen}, \citenamefont {Li}, \citenamefont {Zheng}, \citenamefont
  {Zhang}, \citenamefont {Hou}, \citenamefont {Yin}, \citenamefont {Liu},
  \citenamefont {Peng},\ and\ \citenamefont {Zhang}}]{Cr1.53_topological_Hall}%
  \BibitemOpen
  \bibfield  {author} {\bibinfo {author} {\bibfnamefont {C.}~\bibnamefont
  {Zhang}}, \bibinfo {author} {\bibfnamefont {C.}~\bibnamefont {Liu}}, \bibinfo
  {author} {\bibfnamefont {J.}~\bibnamefont {Zhang}}, \bibinfo {author}
  {\bibfnamefont {Y.}~\bibnamefont {Yuan}}, \bibinfo {author} {\bibfnamefont
  {Y.}~\bibnamefont {Wen}}, \bibinfo {author} {\bibfnamefont {Y.}~\bibnamefont
  {Li}}, \bibinfo {author} {\bibfnamefont {D.}~\bibnamefont {Zheng}}, \bibinfo
  {author} {\bibfnamefont {Q.}~\bibnamefont {Zhang}}, \bibinfo {author}
  {\bibfnamefont {Z.}~\bibnamefont {Hou}}, \bibinfo {author} {\bibfnamefont
  {G.}~\bibnamefont {Yin}}, \bibinfo {author} {\bibfnamefont {K.}~\bibnamefont
  {Liu}}, \bibinfo {author} {\bibfnamefont {Y.}~\bibnamefont {Peng}},\ and\
  \bibinfo {author} {\bibfnamefont {X.-X.}\ \bibnamefont {Zhang}},\ }\bibfield
  {title} {\bibinfo {title} {{Room-Temperature Magnetic Skyrmions and Large
  Topological Hall Effect in Chromium Telluride Engineered by
  Self-Intercalation}},\ }\href
  {https://doi.org/https://doi.org/10.1002/adma.202205967} {\bibfield
  {journal} {\bibinfo  {journal} {Adv. Mater.}\ }\textbf {\bibinfo {volume}
  {35}},\ \bibinfo {pages} {2205967} (\bibinfo {year} {2023})}\BibitemShut
  {NoStop}%
\bibitem [{\citenamefont {Chen}\ \emph {et~al.}(2023)\citenamefont {Chen},
  \citenamefont {Zhu}, \citenamefont {Lin}, \citenamefont {Niu}, \citenamefont
  {Liu}, \citenamefont {Zhuang}, \citenamefont {Zhang}, \citenamefont {Liang},
  \citenamefont {Sun}, \citenamefont {Chen}, \citenamefont {Hu}, \citenamefont
  {Song}, \citenamefont {Zhou}, \citenamefont {Wu}, \citenamefont {Ge},
  \citenamefont {Yang}, \citenamefont {Zhang},\ and\ \citenamefont
  {Wang}}]{Cr5Te6_thin_film}%
  \BibitemOpen
  \bibfield  {author} {\bibinfo {author} {\bibfnamefont {Y.}~\bibnamefont
  {Chen}}, \bibinfo {author} {\bibfnamefont {Y.}~\bibnamefont {Zhu}}, \bibinfo
  {author} {\bibfnamefont {R.}~\bibnamefont {Lin}}, \bibinfo {author}
  {\bibfnamefont {W.}~\bibnamefont {Niu}}, \bibinfo {author} {\bibfnamefont
  {R.}~\bibnamefont {Liu}}, \bibinfo {author} {\bibfnamefont {W.}~\bibnamefont
  {Zhuang}}, \bibinfo {author} {\bibfnamefont {X.}~\bibnamefont {Zhang}},
  \bibinfo {author} {\bibfnamefont {J.}~\bibnamefont {Liang}}, \bibinfo
  {author} {\bibfnamefont {W.}~\bibnamefont {Sun}}, \bibinfo {author}
  {\bibfnamefont {Z.}~\bibnamefont {Chen}}, \bibinfo {author} {\bibfnamefont
  {Y.}~\bibnamefont {Hu}}, \bibinfo {author} {\bibfnamefont {F.}~\bibnamefont
  {Song}}, \bibinfo {author} {\bibfnamefont {J.}~\bibnamefont {Zhou}}, \bibinfo
  {author} {\bibfnamefont {D.}~\bibnamefont {Wu}}, \bibinfo {author}
  {\bibfnamefont {B.}~\bibnamefont {Ge}}, \bibinfo {author} {\bibfnamefont
  {H.}~\bibnamefont {Yang}}, \bibinfo {author} {\bibfnamefont {R.}~\bibnamefont
  {Zhang}},\ and\ \bibinfo {author} {\bibfnamefont {X.}~\bibnamefont {Wang}},\
  }\bibfield  {title} {\bibinfo {title} {{Observation of Colossal Topological
  Hall Effect in Noncoplanar Ferromagnet $\mathrm{Cr}_5\mathrm{Te}_6$ Thin
  Films}},\ }\href {https://doi.org/https://doi.org/10.1002/adfm.202302984}
  {\bibfield  {journal} {\bibinfo  {journal} {Adv. Funct. Mater.}\ }\textbf
  {\bibinfo {volume} {33}},\ \bibinfo {pages} {2302984} (\bibinfo {year}
  {2023})}\BibitemShut {NoStop}%
\bibitem [{\citenamefont {S{\"u}rgers}\ \emph {et~al.}(2014)\citenamefont
  {S{\"u}rgers}, \citenamefont {Fischer}, \citenamefont {Winkel},\ and\
  \citenamefont {L{\"o}hneysen}}]{Mn5Si3}%
  \BibitemOpen
  \bibfield  {author} {\bibinfo {author} {\bibfnamefont {C.}~\bibnamefont
  {S{\"u}rgers}}, \bibinfo {author} {\bibfnamefont {G.}~\bibnamefont
  {Fischer}}, \bibinfo {author} {\bibfnamefont {P.}~\bibnamefont {Winkel}},\
  and\ \bibinfo {author} {\bibfnamefont {H.~v.}\ \bibnamefont
  {L{\"o}hneysen}},\ }\bibfield  {title} {\bibinfo {title} {Large topological
  hall effect in the non-collinear phase of an antiferromagnet},\ }\href@noop
  {} {\bibfield  {journal} {\bibinfo  {journal} {Nat. Commun.}\ }\textbf
  {\bibinfo {volume} {5}},\ \bibinfo {pages} {3400} (\bibinfo {year}
  {2014})}\BibitemShut {NoStop}%
\bibitem [{\citenamefont {Ding}\ \emph {et~al.}(2017)\citenamefont {Ding},
  \citenamefont {Li}, \citenamefont {Xu}, \citenamefont {Wang}, \citenamefont
  {Hou}, \citenamefont {Liu}, \citenamefont {Liu}, \citenamefont {Wu},\ and\
  \citenamefont {Wang}}]{MnNiGa_topological_Hall}%
  \BibitemOpen
  \bibfield  {author} {\bibinfo {author} {\bibfnamefont {B.}~\bibnamefont
  {Ding}}, \bibinfo {author} {\bibfnamefont {Y.}~\bibnamefont {Li}}, \bibinfo
  {author} {\bibfnamefont {G.}~\bibnamefont {Xu}}, \bibinfo {author}
  {\bibfnamefont {Y.}~\bibnamefont {Wang}}, \bibinfo {author} {\bibfnamefont
  {Z.}~\bibnamefont {Hou}}, \bibinfo {author} {\bibfnamefont {E.}~\bibnamefont
  {Liu}}, \bibinfo {author} {\bibfnamefont {Z.}~\bibnamefont {Liu}}, \bibinfo
  {author} {\bibfnamefont {G.}~\bibnamefont {Wu}},\ and\ \bibinfo {author}
  {\bibfnamefont {W.}~\bibnamefont {Wang}},\ }\bibfield  {title} {\bibinfo
  {title} {{Large topological Hall effect in nonchiral hexagonal MnNiGa
  films}},\ }\href {https://doi.org/10.1063/1.4977560} {\bibfield  {journal}
  {\bibinfo  {journal} {Appl. Phys. Lett.}\ }\textbf {\bibinfo {volume}
  {110}},\ \bibinfo {pages} {092404} (\bibinfo {year} {2017})}\BibitemShut
  {NoStop}%
\bibitem [{\citenamefont {Li}\ \emph {et~al.}(2013)\citenamefont {Li},
  \citenamefont {Kanazawa}, \citenamefont {Yu}, \citenamefont {Tsukazaki},
  \citenamefont {Kawasaki}, \citenamefont {Ichikawa}, \citenamefont {Jin},
  \citenamefont {Kagawa},\ and\ \citenamefont
  {Tokura}}]{MnSi_topologicla_Hall}%
  \BibitemOpen
  \bibfield  {author} {\bibinfo {author} {\bibfnamefont {Y.}~\bibnamefont
  {Li}}, \bibinfo {author} {\bibfnamefont {N.}~\bibnamefont {Kanazawa}},
  \bibinfo {author} {\bibfnamefont {X.~Z.}\ \bibnamefont {Yu}}, \bibinfo
  {author} {\bibfnamefont {A.}~\bibnamefont {Tsukazaki}}, \bibinfo {author}
  {\bibfnamefont {M.}~\bibnamefont {Kawasaki}}, \bibinfo {author}
  {\bibfnamefont {M.}~\bibnamefont {Ichikawa}}, \bibinfo {author}
  {\bibfnamefont {X.~F.}\ \bibnamefont {Jin}}, \bibinfo {author} {\bibfnamefont
  {F.}~\bibnamefont {Kagawa}},\ and\ \bibinfo {author} {\bibfnamefont
  {Y.}~\bibnamefont {Tokura}},\ }\bibfield  {title} {\bibinfo {title} {{Robust
  Formation of Skyrmions and Topological Hall Effect Anomaly in Epitaxial Thin
  Films of MnSi}},\ }\href {https://doi.org/10.1103/PhysRevLett.110.117202}
  {\bibfield  {journal} {\bibinfo  {journal} {Phys. Rev. Lett.}\ }\textbf
  {\bibinfo {volume} {110}},\ \bibinfo {pages} {117202} (\bibinfo {year}
  {2013})}\BibitemShut {NoStop}%
\bibitem [{\citenamefont {Shao}\ \emph {et~al.}(2019)\citenamefont {Shao},
  \citenamefont {Liu}, \citenamefont {Yu}, \citenamefont {Kim}, \citenamefont
  {Che}, \citenamefont {Tang}, \citenamefont {He}, \citenamefont {Tserkovnyak},
  \citenamefont {Shi},\ and\ \citenamefont {Wang}}]{TmFeO/Pt_topologicla_Hall}%
  \BibitemOpen
  \bibfield  {author} {\bibinfo {author} {\bibfnamefont {Q.}~\bibnamefont
  {Shao}}, \bibinfo {author} {\bibfnamefont {Y.}~\bibnamefont {Liu}}, \bibinfo
  {author} {\bibfnamefont {G.}~\bibnamefont {Yu}}, \bibinfo {author}
  {\bibfnamefont {S.~K.}\ \bibnamefont {Kim}}, \bibinfo {author} {\bibfnamefont
  {X.}~\bibnamefont {Che}}, \bibinfo {author} {\bibfnamefont {C.}~\bibnamefont
  {Tang}}, \bibinfo {author} {\bibfnamefont {Q.~L.}\ \bibnamefont {He}},
  \bibinfo {author} {\bibfnamefont {Y.}~\bibnamefont {Tserkovnyak}}, \bibinfo
  {author} {\bibfnamefont {J.}~\bibnamefont {Shi}},\ and\ \bibinfo {author}
  {\bibfnamefont {K.~L.}\ \bibnamefont {Wang}},\ }\bibfield  {title} {\bibinfo
  {title} {{Topological Hall effect at above room temperature in
  heterostructures composed of a magnetic insulator and a heavy metal}},\
  }\href {https://doi.org/10.1038/s41928-019-0246-x} {\bibfield  {journal}
  {\bibinfo  {journal} {Nat. Electron.}\ }\textbf {\bibinfo {volume} {2}},\
  \bibinfo {pages} {182} (\bibinfo {year} {2019})}\BibitemShut {NoStop}%
\bibitem [{\citenamefont {Polesya}\ \emph {et~al.}(2010)\citenamefont
  {Polesya}, \citenamefont {Mankovsky}, \citenamefont {Benea}, \citenamefont
  {Ebert},\ and\ \citenamefont {Bensch}}]{CrTe_magnetic_structure}%
  \BibitemOpen
  \bibfield  {author} {\bibinfo {author} {\bibfnamefont {S.}~\bibnamefont
  {Polesya}}, \bibinfo {author} {\bibfnamefont {S.}~\bibnamefont {Mankovsky}},
  \bibinfo {author} {\bibfnamefont {D.}~\bibnamefont {Benea}}, \bibinfo
  {author} {\bibfnamefont {H.}~\bibnamefont {Ebert}},\ and\ \bibinfo {author}
  {\bibfnamefont {W.}~\bibnamefont {Bensch}},\ }\bibfield  {title} {\bibinfo
  {title} {{Finite-temperature magnetism of $\mathrm{Cr}\mathrm{Te}$ and
  $\mathrm{Cr}\mathrm{Se}$}},\ }\href
  {https://doi.org/10.1088/0953-8984/22/15/156002} {\bibfield  {journal}
  {\bibinfo  {journal} {J. Phys. Condens. Matter.}\ }\textbf {\bibinfo {volume}
  {22}},\ \bibinfo {pages} {156002} (\bibinfo {year} {2010})}\BibitemShut
  {NoStop}%
\bibitem [{\citenamefont {Ghimire}\ \emph {et~al.}(2020)\citenamefont
  {Ghimire}, \citenamefont {Dally}, \citenamefont {Poudel}, \citenamefont
  {Jones}, \citenamefont {Michel}, \citenamefont {Magar}, \citenamefont
  {Bleuel}, \citenamefont {McGuire}, \citenamefont {Jiang}, \citenamefont
  {Mitchell}, \citenamefont {Lynn},\ and\ \citenamefont {Mazin}}]{Y166_kagome}%
  \BibitemOpen
  \bibfield  {author} {\bibinfo {author} {\bibfnamefont {N.~J.}\ \bibnamefont
  {Ghimire}}, \bibinfo {author} {\bibfnamefont {R.~L.}\ \bibnamefont {Dally}},
  \bibinfo {author} {\bibfnamefont {L.}~\bibnamefont {Poudel}}, \bibinfo
  {author} {\bibfnamefont {D.~C.}\ \bibnamefont {Jones}}, \bibinfo {author}
  {\bibfnamefont {D.}~\bibnamefont {Michel}}, \bibinfo {author} {\bibfnamefont
  {N.~T.}\ \bibnamefont {Magar}}, \bibinfo {author} {\bibfnamefont
  {M.}~\bibnamefont {Bleuel}}, \bibinfo {author} {\bibfnamefont {M.~A.}\
  \bibnamefont {McGuire}}, \bibinfo {author} {\bibfnamefont {J.~S.}\
  \bibnamefont {Jiang}}, \bibinfo {author} {\bibfnamefont {J.~F.}\ \bibnamefont
  {Mitchell}}, \bibinfo {author} {\bibfnamefont {J.~W.}\ \bibnamefont {Lynn}},\
  and\ \bibinfo {author} {\bibfnamefont {I.~I.}\ \bibnamefont {Mazin}},\
  }\bibfield  {title} {\bibinfo {title} {{Competing magnetic phases and
  fluctuation-driven scalar spin chirality in the kagome metal
  $\mathrm{Y}\mathrm{Mn}_6\mathrm{Sn}_6$}},\ }\href
  {https://doi.org/10.1126/sciadv.abe2680} {\bibfield  {journal} {\bibinfo
  {journal} {Sci. Adv.}\ }\textbf {\bibinfo {volume} {6}},\ \bibinfo {pages}
  {eabe2680} (\bibinfo {year} {2020})}\BibitemShut {NoStop}%
\bibitem [{\citenamefont {Oda}\ \emph {et~al.}(2001)\citenamefont {Oda},
  \citenamefont {Yoshii}, \citenamefont {Yasui}, \citenamefont {Ito},
  \citenamefont {Ido}, \citenamefont {Ohno}, \citenamefont {Kobayashi},\ and\
  \citenamefont {Sato}}]{Cr3Te4_change_AHE_sign}%
  \BibitemOpen
  \bibfield  {author} {\bibinfo {author} {\bibfnamefont {K.}~\bibnamefont
  {Oda}}, \bibinfo {author} {\bibfnamefont {S.}~\bibnamefont {Yoshii}},
  \bibinfo {author} {\bibfnamefont {Y.}~\bibnamefont {Yasui}}, \bibinfo
  {author} {\bibfnamefont {M.}~\bibnamefont {Ito}}, \bibinfo {author}
  {\bibfnamefont {T.}~\bibnamefont {Ido}}, \bibinfo {author} {\bibfnamefont
  {Y.}~\bibnamefont {Ohno}}, \bibinfo {author} {\bibfnamefont {Y.}~\bibnamefont
  {Kobayashi}},\ and\ \bibinfo {author} {\bibfnamefont {M.}~\bibnamefont
  {Sato}},\ }\bibfield  {title} {\bibinfo {title} {{Unusual Anomalous Hall
  Resistivities of $\mathrm{Cu}\mathrm{Cr}$$_2$$\mathrm{S}$$_{4}$,
  $\mathrm{Cu}$$_{0.5}$$\mathrm{Zn}$$_{0.5}$$\mathrm{Cr}$$_2$$\mathrm{Se}$$_{4}$ and
  $\mathrm{Cr}$$_3$$\mathrm{Te}$$_{4}$}},\ }\href
  {https://doi.org/10.1143/JPSJ.70.2999} {\bibfield  {journal} {\bibinfo
  {journal} {J. Phys. Soc. Japan}\ }\textbf {\bibinfo {volume} {70}},\ \bibinfo
  {pages} {2999} (\bibinfo {year} {2001})}\BibitemShut {NoStop}%
\end{thebibliography}
\end{document}